# What water properties are responsible for the physiological temperature interval limits of warm-blooded organisms?

Leonid A. Bulavin [1], Anatoliy I. Fisenko [2], Nikolay P. Malomuzh [3]

[1] *Kiev National University, Department. of Molecular Physics, 2 Academician Glushkov Ave., Kiev, 03680, Ukraine, bulavin221@gmail.com*

[2] *Oncfec, Inc., 250 Lake Street, Suite 909, St. Catharines Ontario, L2R 5Z4, Canada, afisenko@oncfec.com*

[3] *Department of Theoretical Physics, Odessa National University, 2 Dvoryaskaja Street, Odessa, 65026, Ukraine, malomuzhnp@list.ru*

**Abstract:** The weighty evidences of specific transformations of the thermal motion in pure water in the physiological temperature interval (PTI) from $(30 \pm 3)\,^{\circ}\mathrm{C}$ to $(42 \pm 3)\,^{\circ}\mathrm{C}$ for warm-blooded organisms are presented. It is shown that near the right end of the PTI $(42 \pm 3)\,^{\circ}\mathrm{C}$ the crystal-like thermal motion in water transforms to argon-like one (i.e. the dynamic phase transition (DPT) occurs). It is show that the similar transformation takes also place in water-Mioglobin solutions. It is proposed that the DPT takes also place in the intracellular water, where it stimulates the denaturation of proteins. The restriction of the PTI on the left of $(30 \pm 3)\,^{\circ}\mathrm{C}$ is naturally explained by the clusterization of water molecules, which strongly increases when temperature drops. The middle, ($(36 \pm 1)^{0}\mathrm{C}$), of the PTI for warm-blooded organisms is disposed at the minimum of the heat capacity at constant pressure, that forwards to the stability of heat-exchange for bio-cells.

## 1. Introduction

The existence of at least four temperature anomalies in the properties of liquid water occurring near 15, 30, 45 and 60°C were early noted by Drost-Hansen and his coauthors in [1-7]. These anomalies represent some kind of changes in the behavior of different thermodynamic and kinetic characteristics of water and water-protein solutions. It was also proposed that these anomalies reflect some structural transformations occurring in water.

In this paper our main attention is focused on peculiarities of the thermal motion in pure water at $(42 \pm 3)^0 C$, giving us reasons to think about the dynamic phase transition at this temperature. This statement is motivated by 1) the results of the incoherent neutron scattering study of the thermal motion in water; 2) the temperature dependencies of the kinematic shear viscosity and the dielectric relaxation time in water; 3) the surprising behavior of the entropy diameter in a wide temperature interval. Thus, the right end of the physiological temperature interval for warm-blooded is connected with dynamic phase transition from water states with crystal-like thermal motion to ones with argon-like. At the same time the left end of this range, $(30 \pm 3)^0 C$, is connected with the excess of the permissible pH-level and the essential increase of the clusterization in water.

## 2. Dynamic Phase Transition in Pure and Intracellular Water

One of important arguments in the favor of the surprising change of the thermal motion in water near $T_H \approx (42 \pm 3)^0 C$ is given by the behavior of the residence time $\tau_0$ for water molecules. By definition (see [8]), $\tau_0$ is the characteristic time during which a molecule oscillates near some temporary equilibrium position. After this, during the characteristic time $\tau_1$ it displaces to another equilibrium position. If $\tau_0 \gg \tau_1$ we say that the thermal motion has the crystal-like nature. The physical prerequisite for this type of the thermal motion is the existence of the H-bond network in water. At the same time it is natural to expect that $\tau_1$ is close to the simplest molecular time $\tau_f$, which corresponds to free motion of a molecule on distance $a$ of inter-particle spacing: $\tau_f = a / \upsilon_T$, where $\upsilon_T \sim (k_B T / m)^{1/2}$ is the thermal velocity of a molecule.

For the determination of the residence time and the ratio $\tau_0 / \tau_1 \approx \tau_0 / \tau_f$ for water we will use experimental data [9, 10] for the half-width of the diffusion peak in the spectrum of quasi-elastic incoherent neutron scattering. The half-width of the diffusion peak for crystal-like states of water is described by the expression [11, 12]:

$$\gamma_D(\vec{k}^2) \approx D_s \vec{k}^2 - \tau_0 D_s^{(1)2} \vec{k}^4 + \tau_0^2 D_s^{(1)3} \vec{k}^6 + ... \,, \tag{1}$$

where $D_s$ and $D_s^{(1)}$ are the self-diffusion coefficient and one-particle contribution to it correspondingly, $\vec{k}$ is the transferring wave vector for neutrons. The applicability region of the diffusion approximation is limited by wave vectors: $|\vec{k}| \ll 1/a$, where $a$ is the inter-particle spacing.

Fitting the experimental values of the half-width with the help of (1) we can determine all key parameters for crystal-like states of water: $D_s$, $D_s^{(1)}$ and the residence time $\tau_0$. The temperature dependence of $\tau_0/\tau_f$, determined in such a way, is presented in Fig.1.

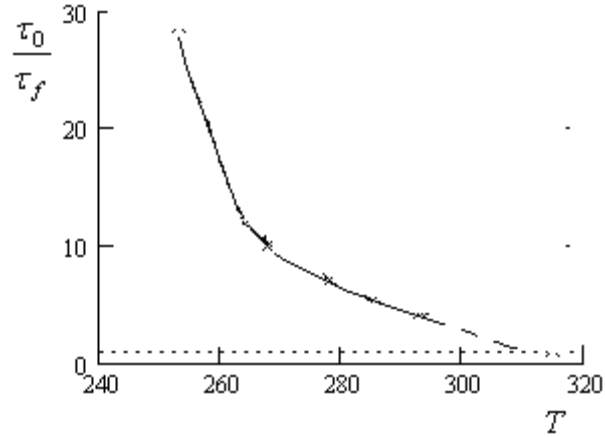

**Figure 1.** The temperature dependence of the ratio $\dfrac{\tau_0}{\tau_f}$, where $\tau_f \approx 5 \cdot 10^{-13}\,s$,

established in [12, 13].

As seen, the inequality $\tau_0/\tau_f > (\gg) 1$, takes only place for super-cooled states of ordinary water and its normal ones within the interval:

$$T_m < T < T_H, \quad T_m = 273\,\text{K}, \quad T_H \approx 315\,\text{K}. \qquad (2)$$

The temperature dependence of the ratio $\tau_0/\tau_f$ allows us to establish the applicability region for crystal-like representations for water. In accordance with Fig.1 we should conclude that crystal-like representations are only applicable for $T < T_H$.

The transition from the crystal-like thermal motion in water to that characteristic for argon-like liquids is also corroborated by the temperature dependence of the kinematic shear viscosity. The last is one of their main transport coefficients. It is formed by different constituents of the thermal motion of molecules in liquids, in the first place, by the translational and rotational degrees of freedom. For water, the considerable influence on their manifestation is produced by H-bonds. Thus, if a molecule is connected with its nearest neighbors by three or four H-bonds, it can only oscillate near some temporary equilibrium position.

For separating contributions of different physical nature, let us compare the behavior of the normalized shear viscosities for water and argon in the manner of the principle of corresponding states [14, 15]. The normalized values of the kinematic shear viscosities for ordinary and heavy water are determined as: $\tilde{v}^{(i)}(t) = v^{(i)}(t)/v_R^{(i)}$ and $\tilde{v}(t) = v^{(Ar)}(t)/v_R^{(Ar)}$. Here $v_R^{(i)}$, $i = H_2O, D_2O$, and $v_R^{(Ar)}$ are the regularized values of the shear viscosities at the critical points [16], $t = T/T_c$ is the dimensionless temperature ($t = T/T_c^{(i)}$ for normal and heavy water and $t = T/T_c^{(Ar)}$ for argon). The temperature dependencies of $\tilde{v}^{(H_2O)}(t)$, $\tilde{v}^{(D_2O)}(t)$ and $\tilde{v}(t)$ are presented in Fig 2.

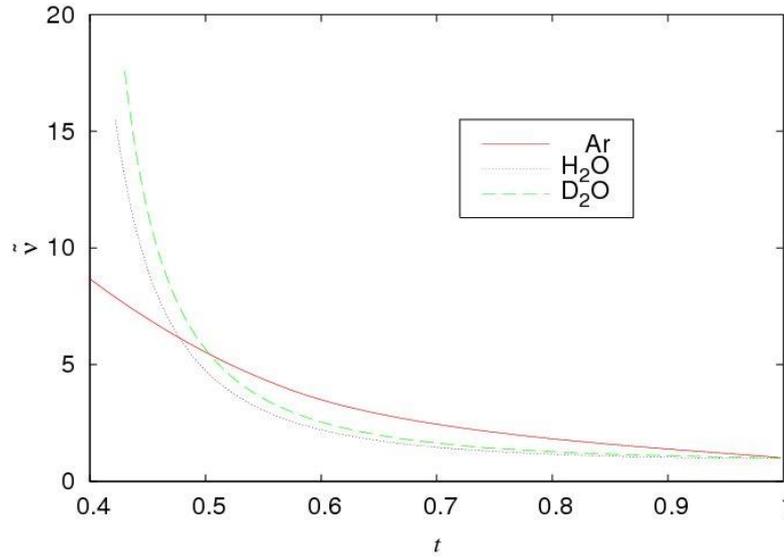

**Figure 2**. Temperature dependencies of the normalized kinematic shear viscosities for argon and the normal and heavy water. The experimental data are taken from. [17]

The points of intersection $t_V$ for the curves $\tilde{v}^{(i)}(t)$, $i = H_2O, D_2O$, and $\tilde{v}(t)$ are the characteristic temperatures for water. It separates two temperature intervals, in which the behavior of the kinematic shear viscosities is determined by essentially different mechanisms. From the equation

$$\tilde{v}^{(i)}(t_V) = \tilde{v}_{ext}(t_V), \qquad (3)$$

where $\tilde{v}_{ext}(t)$ denotes the extrapolated values of the kinematic shear viscosity of argon in its supercooled region, it follows that

$$t_V(H_2O) = 0.478 \Rightarrow T_V(H_2O) = 309 \text{ K},$$

$$t_V(D_2O) = 0.503 \Rightarrow T_V(D_2O) = 324.1 \text{ K}.$$

As we see, the characteristic temperature $T_V(H_2O) = 309$ K for normal water is close to $T_H$. Besides, $T_V$ separates two temperatures ranges with the different character of the temperature dependence of the kinematic shear viscosity.

In [18, 19] it was shown that the kinematic shear viscosity of the normal water $\tilde{\nu}^{(H_2O)}(t)$ for the whole temperature interval of liquid water, including supercooled states and the critical point, can be approximated by the formula:

$$\tilde{\nu}^{(H_2O)}(t) = \lambda \tilde{\nu}(t) + \tilde{\nu}_H(t), \quad \lambda \sim 1. \tag{4}$$

Here $\tilde{\nu}(t)$ is the argon-like contribution and $\tilde{\nu}_H(t)$ is the contribution stimulated by H-bonds. For $T < T_\nu$ the behavior of $\tilde{\nu}_H(t)$ is approximated by exponential function $\tilde{\nu}_H(t) = \tilde{\nu}_0^{(H)} \exp(\varepsilon/t)$, where $\varepsilon = E_a/k_B T_c$, $E_a$ is the activation energy. In [19] it was shown that the dimensionless activation energy $\varepsilon$ coincides with the dimensionless activation energy $\varepsilon_H$ for the dipole relaxation time and is very close to the break energy per H-bond. For $T > T_\nu$ the behavior of $\tilde{\nu}_H(t)$ changes linearly with temperature, i.e. it has nothing of the kind with the exponential dependence (see [19]). At that $|\tilde{\nu}_H(t)|$ is noticeably smaller than argon-like contribution in (4). It means that the shear viscosity of water is formed by argon-like mechanism, which is weakly modified by H-bonds.

Now we complete our observations by the consideration of the entropy diameter

$$S_d^{(w)} = S_\nu^{(w)} + S_l^{(w)} - 2S_c^{(w)},$$

where $S_c^{(w)}$ is the value of entropy at the critical point, which is the more fine thermodynamic characteristics of a system. It characterizes the degree of asymmetry of the vapor and liquid branches of the entropy for water. The behavior of $S_d^{(w)}$ for normal and heavy water ($D_2O$) is presented in Fig. 3 [20]. For the comparison the entropy diameter for argon as well as for the water homologues $H_2S$ and molecular oxygen $O_2$ are also presented in Fig.3. We see that the behavior of $S_d^{(H_2O)}(T)$ is qualitatively different from $S_d^{(Ar)}(T)$. Unlike the latter the temperature dependence of $S_d^{(H_2O)}(T)$ is nonmonotonic and it vanishes in two points. Here we would like to pay attention on the locations of the roots, closed to the melting temperatures of normal and heavy water:

$$T_s^{(H_2O)} = 0.484 \cdot 648.7 = 314.3 \, \text{K} \Rightarrow 41.2^0\text{C}$$

$$T_s^{(D_2O)} = 0.438 \cdot 648.7 = 282.3 \, \text{K} \Rightarrow 9^0\text{C}$$

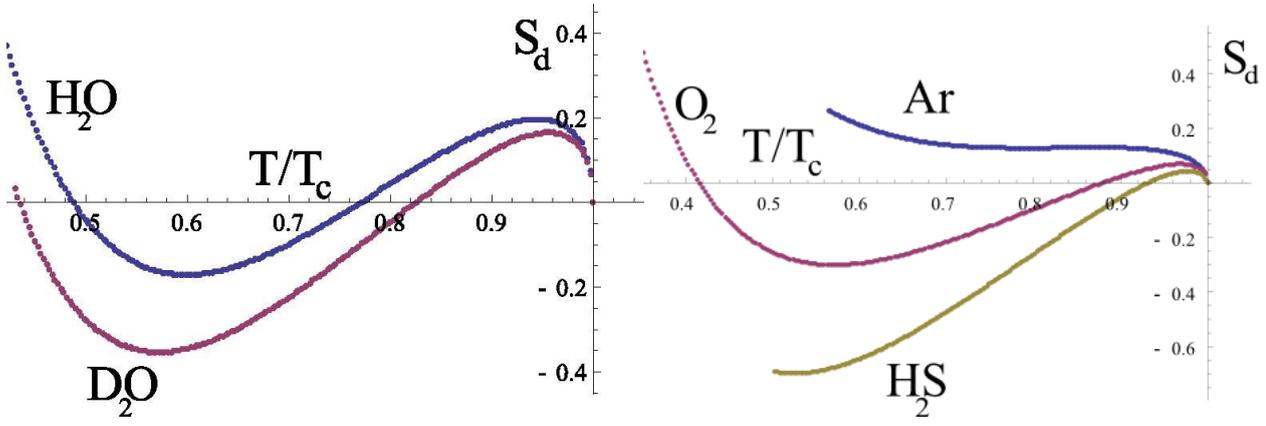

**Figure 3.** The comparative behavior of the entropy diameter for normal and heavy water according to the experimental data [20].

From Fig. 3 it follows (see details in [21]) that qualitatively such a behavior of the entropy diameter is also characteristic for $H_2S$ as well as for simpler systems $N_2$, $O_2$, $F_2$ on their coexistence curves. Therefore we should conclude that the nonmonotonic behavior of $S_d^{(H_2O)}(T)$ is connected with the rotational motion of water molecules. The rapid variation of $S_d^{(H_2O)}(T)$ and $S_d^{(D_2O)}(T)$ near $T_s$ is naturally explained by more essential reduction of the rotational degrees of freedom in the liquid state comparatively with that in vapor one. It is clear that the character of the rotation depends on the number of H-bonds connecting a molecule with its nearest neighbors (details are in [22]).

The direct information about the rotational motion of molecules is given by the temperature dependence of the ratio $\tau_d/\tau_r$, where $\tau_d$ is the dipole relaxation time and $\tau_r$ is the period of the free rotation ($\tau_r \sim \tau_f \approx 0.5 \cdot 10^{-12}$ s).

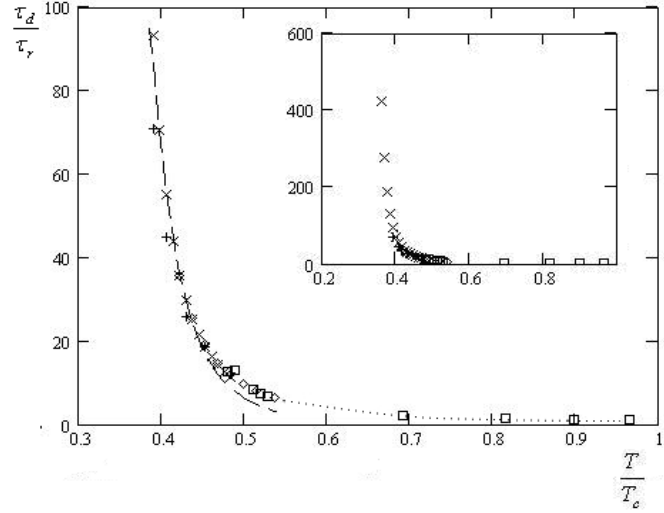

**Figure 4.** The temperature dependencies of the ratio $\tau_d / \tau_r$ for $\tau_d$ taken from different experimental works: crosses – [23], squares – [24], rhombs – [25], pluses – [26], the dashed line corresponds to (5). Points note the interpolation values of $\tau_d$.

As seen from Fig. 4, the noticeable deviation from the exponential dependence (dashed line)

$$\tau_d / \tau_r = 5.1 \cdot 10^{-4} \exp(\varepsilon_H / t), \varepsilon_H \approx 4.71, \qquad (5)$$

is observed near $t_d \approx 0.48$. It is important that the numerical value of the dimensionless activation energy ($\varepsilon_H \approx E_a / k_B T_c$) practically coincides with the H-bonding one. Thus, the change of the rotational motion for water molecules takes place at the same temperature, which is characteristic for the translational motion of molecules.

All characteristic temperatures $T_n, T_v, T_s$ and $T_d$ are close to each other and allows to us to state that near $T_H \approx 315$ K essential change of the thermal motion in water is occurred. In other words the we can speak about the dynamic phase transition in pure water at $T \approx T_H$.

Below we will show that the similar phase transition occurs also in the water-Mioglobin solution. For this purpose we apply to experimental data [27] for the integral intensity $I(\vec{k}^2)$ of incoherent neutron scattering in this solution. In accordance with [27] the mean square displacement of water molecules in water/mioglobin solution is connected with $I(\vec{k}^2)$ by the expression:

$$<\Delta \vec{r}^2>_v = -\frac{3}{2\vec{k}_0^2 \sin^2 \frac{\theta}{2}} \ln\left(\frac{I(\vec{k}^2)}{2\pi}\right), \qquad (6)$$

where $\theta$ is the scattering angle and $\vec{k}_0$ is the wave vector for incident neutron beam. The temperature dependencies of $<\Delta \vec{r}^2>_v$, calculated according to (6) at $|\vec{k}_0| = 1 \text{Å}^{-1}$, is presented in Fig.5.

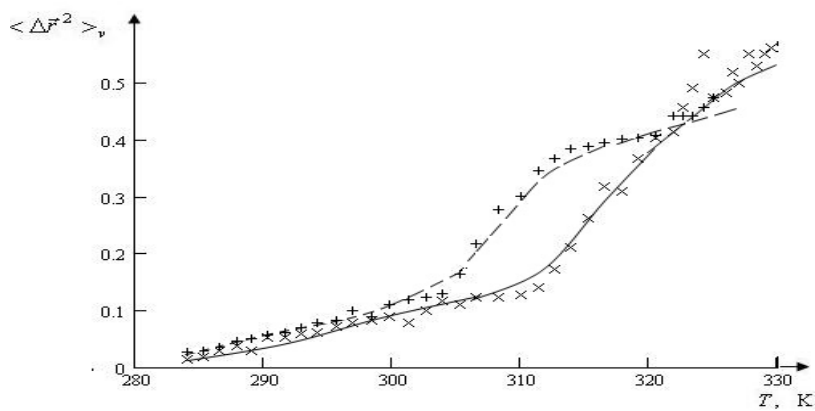

**Figure 5**. The temperature dependence of the MSD of a water molecule for the solution water-Mioglobin

As we see, the MSD of a water molecule increases essentially in the vicinity of the same characteristic temperature $T_H$.

It is very important to note that peculiarities of thermal motion in pure water conserve also for concentrated enough solutions. So the minimum of the compressibility for water-glycerol solution conserve up to 30 mol% [28]). The fine structure of the NMR-spectrum for water-glycerol solution with close concentrations manifests itself only for $T < T_H$ [29].

Now we want to summarize the results obtained in this Section. Analyzing the temperature dependencies of the residence time for water molecules, the shear viscosity, the entropy diameter for pure water and the MSD for water-Mioglobin solution we conclude that the thermal motion in water is undergone to the specific change near $42^0 \text{C}$. This temperature can be interpreted as the temperature of the dynamic phase transition in water and water solutions. It is the upper limit for the crystal-like character of the thermal motion in water. For $T > T_H$ 1) the thermal motion in water tends to the argon-like type and 2) the H-bond network disintegrates in the set of linear multimers (dimers, trimers and so on).

The similar change of the thermal motion in heavy water takes place in the essentially wider temperature interval: $T_v - T_s \approx 51.8 \text{K}$, so we cannot speak about the dynamic phase transition in it.

The closeness of $T_H$ with the upper temperature limit for the life of many warm-blooded organisms allows us to suppose that the dynamic phase transition in the intracellular water stimulates the denaturation of proteins for warm blooded organisms [30, 31]. It seems to us this circumstance is often ignored the in standard approaches to the denaturation problem (see [32]).

## 3. Center of the Physiological Temperature Interval and its Lower Limit

Besides, here it is appropriate to note that the middle of the physiological temperature interval ($(30 \pm 3)^0 C, (42 \pm 3)^0 C$) for warm-blooded organisms practically coincides with the minimum of the heat capacity of water at constant pressure [33] (Fig.6). This minimum is located near $T_{C_p} = (36 \pm 1)^0 C$.

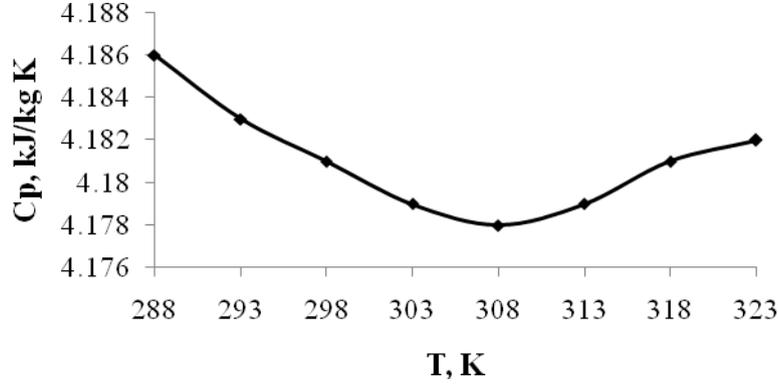

**Figure 6.** Temperature dependence of the heat capacity at constant pressure according to [33].

Such a behavior of the heat capacity ensures the minimal level of the entropy fluctuations: $<(\Delta S)^2> = C_P$, that is very important for the stabile functioning of bio-cells. The value of $C_p \approx 4.1781 \frac{kJ}{kg\ K}$ [17, 33], so the intracellular water strongly influences on the behavior of proteins.

As it noted by Drost-Hansen in [2], many kind of bacteria shows optimum of the physiological activity at 36°C. He also proposed that this temperature was selected as the body temperature for mammals because all physiological processes in them will be most stable to temperature variations of relationship.

In [34], it was shown that the developments of the bacteria studied were suppressed at temperatures 30±2°C and 45±2°C, due to the strong correlation between the structure of water and biological activities. From our point of view the biological activity is tightly connected with the size $r_{cl}$ of water clusters. In [12, 35] it was shown that $r_{cl}$ is naturally identified with the radius $r_L \sim \sqrt{\nu^{(H_2O)}(t)\tau_M}$ of the Lagrange hydrodynamic particle, determining the collective thermal drift of molecules in liquids (here $\nu$ is the kinematic shear viscosity and $\tau_M$ is the Maxwell relaxation time). As it noted in [22], $\tau_M$ is connected with the life time of hexagonal clusters, which determine the short-living local ice-like structure of water. At the same time the corresponding life time is proportional to the well-known dipole relaxation time, so $\tau_M \approx \frac{1}{21}\tau_d$. As a result, we obtain the estimate: $r_{cl}(t) \approx \sqrt{\frac{1}{21}\nu^{(H_2O)}(t)\tau_d(t)}$. Since temperature dependencies of $\tau_d$ and $\nu$ are proportional to each other (see [22]), we can write: $r_{cl}(t) = r_{cl}(t_H) \cdot \frac{\nu^{(H_2O)}(t)}{\nu^{(H_2O)}(t_H)}$. At $42^0C$ the size of a cluster is close to $4\,\text{Å}$ and near $30^0C$ it increases up to $5\,\text{Å}$. From Fig. 2 it follows that $r_{cl}(t)$ strongly increases when temperature dips. The sharp increase in the resistance across a bimolecular phospholipid membrane, taking place between $29^oC$ and $30^oC$ [7, 36], is in qualitative correspondence with essential increase of the shear viscosity, cluster sizes and dipole relaxation time at approaching this temperature.

## 4. Conclusion

The temperature interval favorable for the life of warm-blooded organisms, is limited on the right of $T_{C_p}$ by the temperature of the dynamic phase transition $T_H$. The similar constraint on the left side of $T_{C_p}$ could be connected to the essential amplification of the clusterization in water which does not allow passing the water molecules through a cell membrane. Here it is also desirable to pay attention on the increase of the pH-level of water. The latter is very important independent characteristics of the intracellular water. It changes with temperature according to [31]:

$$pH(T) = 11.314 - 9.396t.$$

It equals ($6.830 \mp 0.044$) at $36^0C$. At $42^0C$ pH of water equals ($6.747 \mp 0.044$), i.e. the increment is $-0.083$. The increase of the pH on the same value in comparison with ($6.830 \pm 0.044$) is observed at $30.3^0C$, that is close to the lower limit for the life of warm-blooded organisms. The corresponding value of pH for the left end of the temperature interval ($6.913 \mp 0.04$) coincides with the upper pH limit for intracellular fluid proposed in [37].

## Acknowledgments

On different stages the question concerned in this work was discussed with Professor J. Barthel, Professor D. Govorun, Professor G. Maisano, Professor S. Magazu, Professor G. Malenkov, Professor E.H. Stanley and many others. We express them our cordial gratitude.